\documentclass[slac_one]{revtex4}
\usepackage{graphicx}
\usepackage{fancyhdr}
\usepackage{amsmath,amssymb}
\newcommand{\etal}{\textit{et al.}}
\newcommand{\pd}{\partial}
\newcommand{\LL}{\mathcal{L}}

\newcommand{\tr}[1]{\mathop{\rm tr}\left\{#1\right\}}

\newcommand{\ii}{\mathrm{i}}

\newcommand{\pp}{{\prime\,2}}

\newcommand{\sw}{s_w}
\newcommand{\cw}{c_w}

\newcommand{\vB}{\mathbf{B}}

\newcommand{\vD}{\mathbf{D}}
\newcommand{\vV}{\mathbf{V}}
\newcommand{\vT}{\mathbf{T}}
\newcommand{\vW}{\mathbf{W}}

\newcommand{\vj}{\mathbf{j}}

\newcommand{\vw}{\mathbf{w}}
\newcommand{\vpi}{\boldsymbol{\pi}}

\newcommand{\vrho}{\boldsymbol{\rho}}

\newcommand{\GeV}{{\ensuremath\rm GeV}}
\newcommand{\TeV}{{\ensuremath\rm TeV}}

\begin{document}

\preprint{DESY 05--115}
\title{{\small{2005 International Linear Collider Workshop - Stanford,
U.S.A.}}\\ 
\vspace{12pt}
Resonances and Electroweak Observables at the ILC} 

%

\author{W. Kilian, J. Reuter}
\affiliation{Deutsches Elektronen-Synchrotron DESY, D--22603 Hamburg, Germany}

\begin{abstract}
Precise measurements of the interactions of electroweak vector bosons
at the ILC yield information about the physics of electroweak symmetry
breaking.  In order to combine this with the possible observation of
new resonances or the effects of new strong interactions at the LHC,
we need to relate resonance parameters with low-energy observables.
We derive these relations for a generic setup and draw conclusions
about the new-physics reach of the ILC in the electroweak sector.
\end{abstract}

\maketitle

\thispagestyle{fancy}


\section{INTRODUCTION} 
The previous generation of precision experiments, in particular data
from LEP and SLC, have established electroweak interactions as a
spontaneously broken gauge theory.  However, the underlying physics
that triggers the formation of a scalar (Higgs) condensate and thus
breaks the symmetry is still unknown.  The possible scenarios range
from purely weakly interacting models, such as the minimal Standard
Model with a light Higgs boson and its supersymmetric generalizations,
to strongly-interacting settings that could indicate the opening-up of
further gauge sectors or extra dimensions.

The electroweak gauge bosons $W$ and $Z$ acquire longitudinal
polarization components via their interaction with the Higgs
condensate.  Thus, a comprehensive study of weak-boson interactions
amounts to a nontrivial measurement of parameters that are related to
the unknown symmetry-breaking sector.  It may happen that this new
physics involves resonances in the elastic scattering of vector bosons
and, in analogy with the form factors of QCD, in the form factors of
vector boson production.  The possible resonances include scalar,
vector or tensor states.  In strongly-interacting models, we expect
new physics to emerge around the scale of unitarity saturation,
$1.2\;\TeV$~\cite{Uni}.  Alternatively, the scattering amplitudes and
form factors might be featureless while saturating the unitarity limit
at high energies.

At the LHC, the energy effectively available for $WW$ scattering
processes may allow for direct observation of these effects, at least
in some of the possible channels~\cite{LHCres}.  The ILC energy is
limited to about $1\;\TeV$, therefore direct observation~\cite{ILCres}
is less likely.  However, the virtual exchange of a massive resonance
has an indirect effect on the measurable interactions of vector
bosons, and therefore is encoded in the values of anomalous couplings
in the low-energy effective theory.  Given the limited amount of data
that will be available at the LHC for these processes, the
complementary measurement of low-energy parameters at the
ILC~\cite{ILClow} is essential for obtaining a complete picture.

The effective theory is given by the chiral Lagrangian formalism for
the (Higgsless) Standard Model~\cite{chL,WWrev}.  The terms in the
Lagrangian are expanded in terms of powers of $1/4\pi v$, where
$v=246\;\GeV$ is the electroweak scale.  The fields are defined in
such a way that an $SU(2)_L$ invariance is manifest.  A corresponding
$SU(2)_R$ invariance, that after spontaneous symmetry breaking would
result in a $SU(2)_{L+R}$ custodial symmetry, is also included, but
explicitly broken down to the hypercharge $U(1)$ subgroup in a
well-defined way.

The Standard Model with a Higgs boson is included in this framework as
a special case, where a scalar resonance is coupled to the electroweak
sector with a specific coupling strength.

In this talk, we present results for the low-energy effects of
resonances that can be coupled to vector bosons.  Given a measurement
of the low-energy parameters (anomalous couplings), this allows one to
relate LHC (high-energy) and ILC (low-energy) observables in the
electroweak sector in a straightforward way.  In~\cite{Krst,WWnew}, this
program is carried out explicitly, using a new determination of the
ILC sensitivity to the anomalous couplings in vector boson
interactions.

\section{THE CHIRAL LAGRANGIAN FRAMEWORK}
The Chiral Lagrangian is a version of the generic low-energy
Lagrangian of electroweak physics that is particularly suited for
studying the symmetry properties of the possible interactions.  The
electroweak field strengths are written in a manifestly $SU(2)_L\times
SU(2)_R$ covariant way:
\begin{align}
  \vW_{\mu\nu} &= \pd_\mu\vW_\nu - \pd_\nu\vW_\mu + ig[\vW_\mu, \vW_\nu],
&
  \vB_{\mu\nu} &= \Sigma\left(\pd_\mu B_\nu - \pd_\nu B_\mu\right)
                  \frac{\tau^3}{2}\Sigma^\dagger,
\end{align}
where $\vW = W^k\tfrac{\tau^k}{2}$.  The symmetry breaking field
$\Sigma$ contains the longitudinal degrees of freedom in terms of the
Goldstone triplet $\vw \equiv w^k\tau^k$,
\begin{equation}
  \Sigma = \exp\left(-\frac{\ii}{v}\vw\right),
\end{equation}
where the exponential parameterization is a convenient one out of
infinitely many equivalent choices.  Furthermore, there is a vector
field~$\vV$ derived from the covariant derivative $\vD\Sigma$ and a
$SU(2)_R$-breaking spurion field~$\vT$,
\begin{align}
  \vV &= \Sigma(\vD\Sigma)^\dagger = -(\vD\Sigma)\Sigma^\dagger,
&
  \vT &= \Sigma\tau^3\Sigma^\dagger.
\end{align}

With these definitions, the basic Lagrangian reads~\cite{chL,WWrev,WWnew}
\begin{equation}\label{eq:chL}
  \LL_0 = -\frac{1}{2}\tr{\vW_{\mu\nu}\vW^{\mu\nu}}
        -\frac{1}{2}\tr{\vB_{\mu\nu}\vB^{\mu\nu}}
        -\frac{v^2}{4}\tr{\vV_\mu\vV^\mu}
	+\beta_1\LL_0'+\sum_i\alpha_i\LL_i.
\end{equation}
The list of lowest-order anomalous couplings is given by the operators
\begin{subequations}
\begin{align}
  \LL_0' &= \tfrac{v^2}{4}\tr{\vT\vV_\mu}\tr{\vT\vV^\mu},
\\
  \LL_1 &= gg' \tr{\vB_{\mu\nu}\vW^{\mu\nu}},
\\
  \LL_2 &= \ii g' \tr{\vB_{\mu\nu}[\vV^\mu,\vV^\nu]},
\\
  \LL_3 &= \ii g \tr{\vW_{\mu\nu}[\vV^\mu,\vV^\nu]},
\\
  \LL_4 &= \left(\tr{\vV_\mu\vV_\nu}\right)^2,
\\
  \LL_5 &= \left(\tr{\vV_\mu\vV^\mu}\right)^2,
\\
  \LL_6 &= \tr{\vV_\mu\vV_\nu} \tr{\vT\vV^\mu} \tr{\vT\vV^\nu},
\\
  \LL_7 &= \tr{\vV_\mu\vV^\mu} \left(\tr{\vT\vV_\nu}\right)^2,
\\
  \LL_8 &= \tfrac14 g^2 \left(\tr{\vT\vW_{\mu\nu}}\right)^2,
\\
  \LL_9 &= \tfrac12 \ii g
             \tr{\vT\vW_{\mu\nu}} \tr{\vT[\vV^\mu,\vV^\nu]},
\\
  \LL_{10} &= \tfrac12
             \left(\tr{\vT\vV_\mu}\right)^2 \left(\tr{\vT\vV_\nu}\right)^2.
\end{align}
\end{subequations}
In this list, there are three operators ($\LL_0',\LL_1,\LL_8$) that
affect gauge-boson propagators directly (oblique corrections).  They
are related to the $S,T,U$ parameters~\cite{STU}.  Three additional
operators ($\LL_2,\LL_3,\LL_9$) contribute to anomalous triple gauge
couplings~\cite{TGC}.  The remaining five operators
($\LL_{4}$--$\LL_{7}$ and $\LL_{10}$) induce anomalous quartic
couplings only.

For the purpose of this talk, we have restricted ourselves to the
purely bosonic part of the effective Lagrangian.  This is appropriate
if the new physics is not directly coupled to the fermions of the
Standard Model.  The extension to fermionic couplings is
straightforward; we have to add all kinds of four-fermion operators
together with a list of operators that couple bilinear fermion
currents to bosonic currents and thus affect the effectively measured
$W$ and $Z$ couplings to fermions.  Furthermore, operators that
violate C and/or CP can be added as well; we do not include these in
the present work.

\section{RESONANCES}
If we do not consider direct couplings to SM fermions, the properties
of new resonances are encoded in their couplings to the bosonic
degrees of freedom.  These terms have to be added to the
Lagrangian~(\ref{eq:chL}).  The couplings are organized in powers of
$1/M$, where $M$ is the resonance mass.  If $M\gg v$ (or if, in the
light-Higgs case, we want to have a renormalizable model), we can
truncate the series after the leading term(s), so only a finite number
of parameters is involved.

A further organizing principle is the number of spurions $\vT$ that
have to be introduced.  Without these, i.e., in the custodial
symmetric (or isospin-conserving) case, only certain combinations are
allowed: a resonance coupled to $WW/ZZ/WZ$ pairs must have the $(I,J)$
[isospin,spin] assignments $(0,0)$, $(2,0)$, $(0,2)$, $(2,2)$, or
$(1,1)$.  However, the presence of the isospin-breaking coupling of
the $B$ field in the leading-order Lagrangian precludes exact isospin
conservation.  Allowing for breaking terms, we have to consider all
isospin/spin assignments of resonances up to~$2$.  In this talk, we
omit the spin-2 and isospin-2 states for brevity; for a full account,
see~\cite{WWnew}.

For each resonance type, we write down the extra terms in the
Lagrangian.  A direct consequence is the result for the two-boson
partial decay widths, which we compute in the limit $M\gg M_W,M_Z$
where the Goldstone-boson equivalence theorem can be used.

In the ILC context, we are interested in the low-energy effects.
Therefore, we also compute the result from integrating out the
resonance at tree level, retaining terms up to order $(v/M)^4$.  This
result is expressed in values for the anomalous couplings $\beta_1$
and $\alpha_i$.  Note that other sources of anomalous couplings,
namely loop contributions within the Standard Model and beyond, are
not given here.

\subsection{Scalar Resonances}
\label{sec:scalar}
Scalar resonances are of particular interest since the most prominent
representative of this class, a $I=0$ scalar boson, serves as a Higgs
boson if its couplings take particular values.  In extended models
with Higgs bosons, there are also scalar resonances with higher
isospin.  For instance, in the MSSM the $(H^+,H^0,H^-)$ triplet can be
viewed as an $I=1$ triplet.  As another example, some Little Higgs
models contain a complex triplet $(\phi^{++},\phi^+,\phi^0)$, which
under isospin decomposes into a real $I=2$ quintet and a singlet.

\subsubsection{Scalar Singlet: \boldmath$\sigma$}
This state has two independent linear couplings, $g_\sigma$ and
$h_\sigma$.  The latter violates isospin.  Neglecting self-couplings
etc.\ that do not contribute to the order we are interested in, the
Lagrangian is
\begin{equation}
  \LL_\sigma = -\frac12\left[\sigma\left(M_\sigma^2 + \pd^2\right)\sigma
                             + 2\sigma j\right],
\qquad\text{where}\qquad
  j = -\frac{g_\sigma v}{2}\tr{\vV_\mu\vV^\mu}
      -\frac{h_\sigma v}{2}\left(\tr{\vT\vV_\mu}\right)^2.
\end{equation}
The Higgs boson corresponds to the special values $g_\sigma=1$ and
$h_\sigma=0$.  Integrating out $\sigma$, we get zero shifts for
all anomalous couplings, except for
\def\fact{\left(\frac{v^2}{8M_\sigma^2}\right)}
\begin{align}
  \alpha_5 &= g_\sigma^2\fact,
&
  \alpha_7 &= 2 g_\sigma h_\sigma\fact,
&
  \alpha_{10} &=  2h_\sigma^2\fact.
\end{align}
In the high-mass limit, the $\sigma$ width is given by
\begin{equation}
  \Gamma_\sigma = \frac{g_\sigma^2 + \frac12(g_\sigma+2h_\sigma)^2}{16\pi}
                  \left(\frac{M_\sigma^3}{v^2}\right).
\end{equation}
This includes $\sigma\to W^+W^-$ and $\sigma\to ZZ$.

\subsubsection{Scalar Triplet: \boldmath$\pi$}
Coupling a scalar triplet $\vpi$ to vector boson pairs
requires isospin violation.  The Lagrangian is
\begin{equation}
  \LL_\pi = -\frac14\tr{\vpi(M_\pi^2 + \vD^2)\vpi + 2\vpi\vj}
\quad\text{with}\quad
  \vj = \frac{h_\pi v}{2}\vV_\mu\tr{\vT\vV^\mu}
       +\frac{h_\pi'v}{2}\vT\tr{\vV_\mu\vV^\mu}
       +\frac{k_\pi v}{2}\vT\left(\tr{\vT\vV_\mu}\right)^2.
\end{equation}
Evaluating the effective Lagrangian, the nonvanishing parameters are
\def\fact{\left(\frac{v^2}{16M_\pi^2}\right)}
\begin{align}
  \alpha_5 &= 2h_\pi^\pp\fact,
&
  \alpha_6 &= h_\pi^2\fact,
&
  \alpha_7 &= 2h_\pi'(h_\pi+ 2k_\pi)\fact,
&
  \alpha_{10} &=  4k_\pi(h_\pi + k_\pi)\fact.
\end{align}
The partial widths for the decay into vector boson pairs are different
for charged and neutral pions:
\begin{subequations}
\begin{align}
  \Gamma_{\pi^\pm} &=
  \frac{\frac14h_\pi^2}{16\pi}\left(\frac{M_\pi^3}{v^2}\right),
&
  \Gamma_{\pi^0} &= 
  \frac{h_\pi^\pp + \frac12(h_\pi + h_\pi' + 2k_\pi)^2}{16\pi}
  \left(\frac{M_\pi^3}{v^2}\right).
\end{align}
\end{subequations}
If there is approximate isospin conservation we expect the total
widths to be dominated by fermion pairs and by three-boson decays,
analogous to the pions of QCD.

\subsection{Vector Resonances}
\label{sec:vector}
In contrast to scalar resonances, vector resonances contribute to
operators involving field strengths and thus contribute to two- and
three-gauge boson couplings.  For that reason, and because in the
isospin-conserving case the leading contribution vanishes, we expand
up to order $1/M^4$.

\subsubsection{Vector singlet: $\omega$}
The Lagrangian is
\begin{equation}
  \LL_\omega =
  \frac12\left[\omega_\mu\left((M^2+\pd^2)g^{\mu\nu}-\pd^\nu\pd^\mu\right)
    \omega_\nu + 2\omega_\mu j^\mu\right]
\qquad\text{with}\qquad
  j_\mu = \ii\frac{h_\omega v^2}{2}\tr{\vT\vV_\mu}
  + \ii\frac{2\ell_\omega}{M_\omega^2}\pd_\nu
    \tr{\vT\vW^\nu{}_\rho\vW^\rho{}_\mu}.
\end{equation}
Expanding up to second order and expressing the result in the
canonical operator basis, we obtain the nonvanishing coefficients
\def\fact{\left(\frac{v^2}{2M_\omega^2}\right)}
\begin{subequations}
\begin{gather}\label{eq:b1om}
  \beta_1 = h_\omega^2 \frac{v^2}{2M_\omega^2},
\qquad
  \alpha_1 = \alpha_2 = \alpha_4 = \alpha_7 = h_\omega^2 \fact^2,
\qquad
  \alpha_5 = \alpha_6 = \alpha_8 = \alpha_9 = -h_\omega^2 \fact^2.
\end{gather}
\end{subequations}
The $\omega$ boson can decay into $W^+W^-$ but not into $ZZ$, and the
pair decay width is
\begin{equation}
  \Gamma_\omega = \frac{h_\omega^2 + \frac12\ell_\omega^2}{48\pi}M_\omega.
\end{equation}

\subsubsection{Vector triplet: $\rho$}
The Lagrangian for a vector triplet $\rho$ contains anomalous magnetic
moments and takes the form
\begin{align}
  \LL_\rho = \frac14\tr{\vrho_\mu\left(
    M_\rho^2 g^{\mu\nu} + \vD^2 g^{\mu\nu} - \vD^\nu\vD^\mu 
    + 2\ii\mu_\rho g\vW^{\mu\nu} + 2\ii\mu_\rho' g'\vB^{\mu\nu}\right)
    \vrho_\nu
    + 2\vrho_\mu\vj^\mu},
\end{align}
where
\begin{align}
  \vj_\mu &= \ii g_\rho v^2\vV_\mu + \ii g_\rho'v^2\vT\tr{\vT\vV_\mu}
  + \ii\frac{4\ell_\rho}{M_\rho^2}\vD_\nu
    \left(\vW^\nu{}_\rho\vW^{\rho\mu}\right)
  + \ii\frac{4\ell_\rho'}{M_\rho^2}\vD_\nu
    \left(\vB^\nu{}_\rho\vW^{\rho\mu}\right)
  + \ii\frac{4\ell_\rho''}{M_\rho^2}\vD_\nu
    \left(\vT\tr{\vT\vW^\nu{}_\rho\vW^{\rho\mu}}\right).
\end{align}
Including a mass splitting $\Delta M_\rho$, we get the nonvanishing
parameters 
\def\fact{\left(\frac{v^2}{2M_\rho^2}\right)}
\begin{subequations}
\begin{align}\label{eq:b1rho}
  \beta_1 &= 4 h_\rho(g_\rho + h_\rho)\frac{v^2}{2M_\rho^2}
  - (g_\rho + 2h_\rho)^2\frac{v^2\Delta M_\rho^2}{2M_\rho^4},
  &
  \alpha_1 &= (g_\rho + 2h_\rho)^2\fact^2,
\\
  \alpha_2 &= \left(-g_\rho^2 + 4h_\rho^2 
                    + g_\rho^2{\mu_\rho'}\right)\fact^2,
& 
  \alpha_3 &= g_\rho(g_\rho+2h_\rho)
              \left(1 + {\mu_\rho} \right)\fact^2,
\\
  \alpha_4 &= (g_\rho - 2h_\rho)^2\fact^2,
&
  \alpha_5 &= -(g_\rho - 2h_\rho)^2\fact^2,
\\
  \alpha_6 &= 8g_\rho h_\rho\fact^2,
&
  \alpha_7 &= -8g_\rho h_\rho\fact^2,
\\
  \alpha_8 &= -4h_\rho(g_\rho+h_\rho)\fact^2,
&
  \alpha_9 &= -2\left[h_\rho(g_\rho+2h_\rho) 
                     + g_\rho h_\rho{\mu_\rho}\right]\fact^2.
\end{align}
\end{subequations}
These parameters modify the triple gauge couplings $g_1^Z$,
$\kappa_\gamma$ and $\kappa_Z$, along with oblique corrections and
quartic gauge couplings.  At the same order of the expansion, there
are also contributions to $\lambda_\gamma,\lambda_Z$ that correspond
to dimension-6 operators not listed here (see~\cite{WWnew}).  As a
particular result, we note that the relation
$\lambda_\gamma=\lambda_Z$ is violated by these terms.

A charged $\rho$ resonance can decay into $W^\pm Z$ and $W^\pm\gamma$.
For the neutral $\rho$, the Landau-Yang theorem forbids $ZZ$ and
$\gamma\gamma$ final states.  The partial widths are
\begin{subequations}
\begin{align}
  \Gamma_{\rho^\pm \to W^\pm Z} &= 
  \frac{(g_\rho + 2h_\rho)^2 
        + 2(\cw\ell_\rho + \frac12\sw\ell_\rho')^2}{48\pi} M_\rho,
&
  \Gamma_{\rho^\pm} &= 
  \frac{(g_\rho + 2h_\rho)^2 
        + \frac12\ell_\rho^2 + \frac1{8}\ell_\rho^\pp}{48\pi} M_\rho,
\\
  \Gamma_{\rho^\pm \to W^\pm\gamma} &=
  \frac{2(\sw\ell_\rho - \frac12\cw\ell_\rho')^2}{48\pi} M_\rho,
&
  \Gamma_{\rho^0} &= 
  \frac{(g_\rho - 2h_\rho)^2
        + 2(\ell_\rho + \frac12\ell_\rho'')^2}{48\pi} M_\rho.
\end{align}
\end{subequations}

\section{CONCLUSIONS}
The low-energy effect of a resonance is proportional to its couplings
squared.  The same holds for the partial width for the decay into
vector boson pairs.  The total width is bounded from below by the
partial widths, while an upper limit is given by the resonance mass:
if the would-be width becomes larger than that, the resonance
parameterization loses its meaning.  Combining these two observations,
we see that the maximally allowed coupling value, and thus the
strongest possible low-energy effect, is present if the partial width
is equal to the mass, and there are no other decay channels.

We conclude that for each given channel and coupling parameter, we
obtain an inequality of the form
\begin{equation}
  |\Delta\alpha_i| \lesssim 4\pi k_i\left(\frac{v^2}{M^2}\right)^2,
\end{equation}
where $k_i$ is some numerical factor.  In the scalar and tensor cases,
the width is proportional to $M^3$, while the leading effect on
anomalous couplings is proportional to $1/M^2$.  In the vector case,
the width scales with the mass, but all anomalous couplings except for
$\beta_1$ get contributions only at order $1/M^4$.  The net effect is
of order $1/M^4$ for (almost) all channels and couplings.  This
clearly limits the physics reach of low-energy measurements.

The significant exception is the $\beta_1$ parameter, which is
equivalent to the $\rho$ parameter.  If isospin is violated, this
parameter receives contributions of the order $1/M^2$,
cf.~(\ref{eq:b1om}, \ref{eq:b1rho}).  We thus confirm the expectation
that this parameter is most sensitive to new-physics effects.
However, we also see that there are ways to suppress this
contribution: (i) If vector-boson interactions conserve isospin
($h_\rho=0$, $\Delta M_\rho=0$), the parameter is zero.
(ii) For the triplet case, there is also a pseudo-symmetric limit
$h_\rho=-g_\rho$, $\Delta M_\rho=0$ where $\beta_1$ vanishes.  (iii)
There may be some cancellation with the mass splitting, or with extra
terms not accounted for by tree-level resonance exchange.  In
particular, we have to keep in mind that fermionic couplings and loop
effects could also be relevant.

It is interesting to note that, at order $1/M^4$, several different
couplings of vector resonances are present and affect low-energy
observables.  In particular, there are contributions of anomalous
magnetic moments.  These cannot be measured by scanning resonance
curves at LHC or at a higher-energy hadron or lepton collider.  Thus,
the precision measurement capabilities of the ILC add independent
information, regardless of the state of LHC high-energy measurements.

\begin{acknowledgments}
This work is supported by the German Helmholtz-Gemeinschaft, Contract
No.\ VH--NG--005.
\end{acknowledgments}


\end{document}